\documentclass[aps,twocolumn,showpacs,preprintnumbers,amsmath,amssymb,superscriptaddress,floatfix,nofootinbib]{revtex4}

\usepackage{graphicx}
\usepackage{epsfig}
\usepackage{epstopdf}
\usepackage{subfigure}

\usepackage{amsmath}
\usepackage{amsfonts}
\usepackage{amssymb}
\usepackage{color}
\usepackage[colorlinks,
            citecolor=blue,
            anchorcolor=red,
            menucolor=red,
            linkcolor=red,
            filecolor=red,
            runcolor=red,
            urlcolor=blue,
            frenchlinks=red]{hyperref}

\begin{document}

\title{Roles of $a_0(980)$, $\Lambda(1670)$, and $\Sigma(1385)$ in the $\Lambda_c^+ \to \eta \Lambda \pi^+$ decay}

\author{Guan-Ying Wang}
\affiliation{Joint Research Center for Theoretical Physics, School of Physics and Electronics, Henan University, Kaifeng 475004, China}

\author{Neng-Chang Wei}
\affiliation{School of Nuclear
Sciences and Technology, University of Chinese Academy of Sciences,
Beijing 101408, China}

\author{Hui-Min Yang}
\affiliation{School of Physics, Beihang University, Beijing 102206,
China}

\author{En Wang}\email{wangen@zzu.edu.cn}
\affiliation{School of Physics and Microelectronics, Zhengzhou
University, Zhengzhou, Henan 450001, China}

\author{Li-Sheng Geng} \email{lisheng.geng@buaa.edu.cn}
\affiliation{School of Physics, Beihang University, Beijing 102206,
China} \affiliation{Beijing Key Laboratory of Advanced Nuclear
Materials and Physics, Beihang University, Beijing 102206, China}
\affiliation{School of Physics and Microelectronics, Zhengzhou
University, Zhengzhou, Henan 450001, China}
\author{Ju-Jun Xie} \email{xiejujun@impcas.ac.cn}
\affiliation{Institute of Modern Physics, Chinese Academy of
Sciences, Lanzhou 730000, China} \affiliation{School of Nuclear
Sciences and Technology, University of Chinese Academy of Sciences,
Beijing 101408, China} \affiliation{School of Physics and
Microelectronics, Zhengzhou University, Zhengzhou, Henan 450001,
China}

\date{\today}

\begin{abstract}
Recently, the Belle Collaboration has measured the $\Lambda_c^+ \to \eta \Lambda \pi^+$ decay and reported the $\eta\Lambda$ and $\Lambda\pi^+ $ invariant mass distributions, which show the clear signals of the resonances $\Lambda(1670)$ and $\Sigma(1385)$, respectively. Based on our previous works [Eur. Phys. J. C 76 (2016) 496 and Phys. Rev. D 95 (2017) 074024], we re-analyze this process by considering the $S$-wave $\eta\Lambda$ and $\eta\pi^+$ final state interactions within the chiral unitary approach, which dynamically generate the $\Lambda(1670)$ and $a_0(980)$, respectively. Our results are in good agreement with the Belle measurements, which supports the molecular nature of the $\Lambda(1670)$ and $a_0(980)$. In addition, the $\eta \pi^+$ invariant mass distributions are also computed and a cusp structure of $a_0(980)$ is cleary shown around the $K\bar{K}$ mass threshold.

\end{abstract}


\maketitle

\section{Introduction}

Compared to the two-body hadronic decays, multi-body hadronic decays of heavy hadrons could be used to explore the nature of the intermediate resonances,  since those processes have much larger phase space, and involve the strong interactions of the final states~\cite{Oset:2016lyh}. In the last decades, many hadron resonances are discovered  in multi-body decays of heavy hadrons by different experiments, such as the $X(3872)$, $P_c$, $P_{cs}$, and $T_{cc}$~\cite{Belle:2003nnu,LHCb:2015yax,LHCb:2019kea,LHCb:2020jpq,LHCb:2017iph,ParticleDataGroup:2020ssz}. The baryon $\Lambda_c$, as the first observed charmed baryon~\cite{Knapp:1976qw}, has a large number of decay modes reported by the BESIII, Belle, {\it BABAR}, CLEO, and LHCb Collaborations~\cite{ParticleDataGroup:2020ssz,Abe:2001mb,Ablikim:2015flg,Pal:2017ypp,Zupanc:2013iki,Aaij:2017rin}, which offers an excellent laboratory for testing the theoretical predictions of the light mesons and  light baryons~\cite{Cheng:2015iom,Klempt:2009pi,Dai:2018hqb,Xie:2018gbi,Hyodo:2011js,Lu:2016ogy,Wang:2020pem,Li:2020fqp,Xie:2017mbe,Xie:2017erh}, such as  the processes $\Lambda^+_c \to p K^+K^-,\, p\pi^+\pi^-$~\cite{Wang:2020pem}, $\Lambda_c^+\to p\eta\pi^0$~\cite{Li:2020fqp}, $\Lambda_c^+ \to \pi^0 \phi p$ decay~\cite{Xie:2017mbe}, and 
$\Lambda_c \to \bar{K}^0 \eta p$ decay~\cite{Xie:2017erh}.

The process of $\Lambda_c^+ \to \Lambda \eta \pi^+$ has  attracted much attentions from  theoreticians and experimentalists, since it provides the fruitful information about the intermediate states.
In Ref.~\cite{Xie:2016evi}, it is proposed to study the $a_0(980)$ and $\Lambda(1670)$ resonances in the $\Lambda_c^+ \to \Lambda \eta \pi^+$ decay via the final-state interactions of the $\eta \pi^+$ and $\Lambda \eta$ pairs. In addition, the process $\Lambda^+_c \to \Lambda \eta \pi^+$ is used to study the resonances $\Lambda(1405)$ and $\Lambda(1670)$ in the $\Lambda \eta$ final state~\cite{Miyahara:2015cja}. Reference~\cite{Xie:2017xwx} has also suggested that this process could be used to search for the missing baryon $\Sigma^*(1/2^-)$ which is interesting theoretically~\cite{Roca:2013cca,Garcia-Recio:2002yxy,Wu:2009nw,Xie:2018gbi,Liu:2017hdx,Wang:2015qta,Xie:2014zga}.

Experimentally, the decay $\Lambda_c^+ \to \Lambda \eta \pi^+$ has been measured by the CLEO Collaboration in 1995~\cite{Ammar:1995je} and 2003~\cite{CLEO:2002aec}. Later in 2009, the BESIII Collaboration has presented an improved measurement of the absolute BFs of the $\Lambda_c^+ \to \Lambda \eta \pi^+$ and studied the intermediate state $\Sigma(1385)^+$ in the three-body decay~\cite{BESIII:2018qyg}. Recently, the Belle Collaboration has also measured the process $\Lambda_c^+ \to \eta \Lambda \pi^+$, and presented the $\eta\Lambda$ and $\Lambda\pi^+$ invariant mass distributions, which show the significant signals of the $\Lambda(1670)$ and $\Sigma(1385)^+$ resonances~\cite{Lee:2020xoz}. Furthermore, the Belle Collaboration has also reported the contribution of $a_0(980)^+$ to the $\eta \pi^+$.

Both the $\Lambda(1670)$ and $a_0(980)$ have generated a lot of interests in their nature. 
For the baryon $\Lambda(1670)$, although the three-quark explanation is supported by the analysis of the high precision data on $K^-p\to \eta \Lambda$~\cite{Starostin:2001zz}, and the study of $K^-p\to \pi^0\Sigma^0$ within the chiral quark model~\cite{Zhong:2008km}, the $S$-wave meson-baryon interactions within the chiarl unitary approach can dynamically generate the resonance $\Lambda(1670)$~\cite{Oset:2001cn,Garcia-Recio:2002yxy,Oller:2006jw}. Indeed, the actual shape of the $\Lambda(1670)$ strongly depends on the process. For instance, the $\Lambda(1670)$ peak is shifted to about 1.7~GeV in the  modulus squared of the $\bar{K}N$ or $\pi\Sigma$ elastic  scattering amplitudes with isospin $I=1$, due to the strong distortion induced by the $\eta\Lambda$ channel, and its shape appears as a clear strong enhancement, not the symmetric Breit-Wigner structure, in the $\eta\Lambda$ mass distribution~\cite{Oller:2006jw}, which is confirmed by the Belle measurements~\cite{Lee:2020xoz}.  In addition, the light scalar meson $a_0(980)$ also has been explained to be either a molecular state, a tetraquark state, a conventional $q\bar{q}$ meson, or the mixing of different components~\cite{Klempt:2007cp,Nieves:1998hp,Janssen:1994wn,Wolkanowski:2015lsa}. In the chiral unitary approach, $a_0(980)$ could be dynamically generated from the $S$-wave interactions of the coupled channels $K\bar{K}$ and $\pi\eta$~\cite{Oller:1997ti,Nieves:1998hp}, which could be used to successfully interpret many experimental measurements~\cite{Wang:2020pem,Xie:2014tma,Oset:2016lyh}.

In this work, we reanalyze the $\Lambda_c^+ \to \Lambda \eta \pi^+$ decay by taking into account the $\Sigma(1385)^+$ decaying into $\pi^+ \Lambda$ in $P$-wave. The dynamical generated states $\Lambda(1670)$ and $a_0(980)$ are also considered. Within the chiral unitary approach, we calculate the invariant mass distributions for $\Lambda_c^+ \to \Lambda \eta \pi^+$ decay. It is found that the experimental measurements of Belle collaboration~\cite{Lee:2020xoz} can be well reproduced.

This article is organized as follows. In Sec.~\ref{Sec:Formalism}, we present the theoretical formalism of the $\Lambda_c^+ \to \eta \Lambda\pi^+ $ decay, and numerical results and discussions are presented in Sec.~\ref{Sec:Results}, followed by a short summary in the last section.

\section{Formalism} \label{Sec:Formalism}

 In this section, we will introduce the theoretical formalism for the process $\Lambda_c^+ \to  \eta \Lambda\pi^+$.  We first  present the mechanism of the intermediate states $\Lambda(1670)$ and $a_0(980)$ production in Subsec.~\ref{sec:molecule}. We will also consider the $\Sigma(1385)$ contribution in the final state interaction of  $\pi\Lambda$ system in Subsec.~\ref{sec:1385}. Finally, we give the formalism of the invariant mass distributions for the process  $\Lambda_c^+ \to  \eta \Lambda\pi^+$ in Subsec.~\ref{sec:width}.

\subsection{Contributions of  $\Lambda(1670)$ and $a_0(980)$}\label{sec:molecule}
\begin{figure}[htbp]
\begin{center}
\includegraphics[scale=0.8]{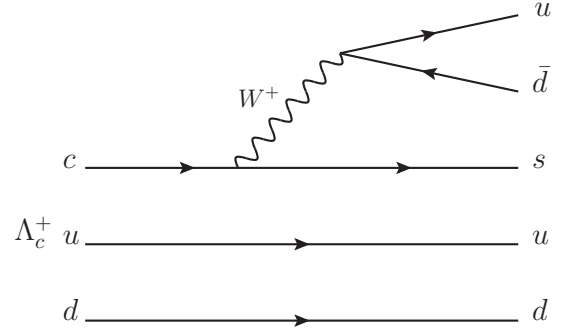}
\caption{Feynman diagram at the quark level for the weakly process $\Lambda^+_c \to W^+ + s + ud\to u\bar{d}+ s + ud$.}
\label{Fig:feynd}
\end{center}
\end{figure}

Analogous to Refs.~\cite{Miyahara:2015cja,Xie:2016evi,Xie:2017xwx}, the process $\Lambda_c^+ \to  \eta \Lambda\pi^+$   proceeds in the following three steps:

1) Weak decay: the $c$ quark of $\Lambda^+_c$
weakly decays into an $s$ quark plus a $W^+$ boson, then the $W^+$ boson goes into a $u \bar d$ pair, as shown in Fig.~\ref{Fig:feynd}.

\begin{figure}[htbp]
\begin{center}
\includegraphics[scale=0.8]{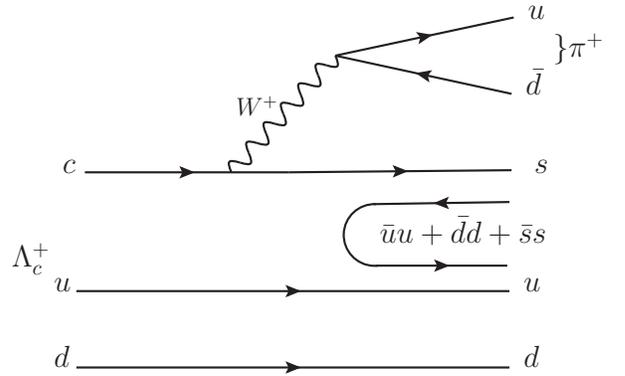}
\caption{Hadronization of the process $\Lambda_c^+ \to \pi^+ M B$. The $u \bar d$ pair from the $W^+$ decay will hadronize into $\pi^+$, and the quark $s$ from the $c$ decay, and the $ud$ pair of the initial $\Lambda$, together with the $\bar{q}q=\bar u u + \bar d d + \bar s s$ created from vacuum, hadronize into a pseudoscalar meson and a baryon,} \label{Fig:mbproduction}
\end{center}
\end{figure}

\begin{figure}[htbp]
\begin{center}
\includegraphics[scale=0.8]{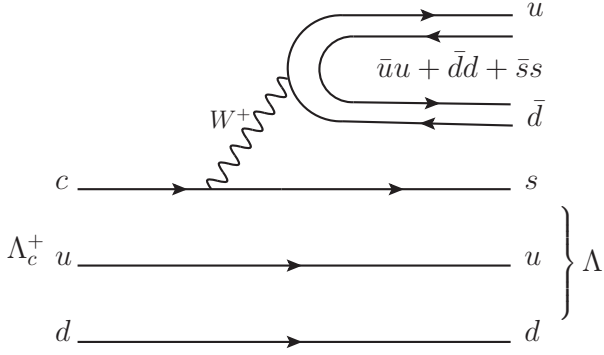}
\caption{Hadronization of the process $\Lambda^+_c \to M M \Lambda$. The $u\bar{d}$ pair from the $W^+$ decay could hadronize into a pair of pseudoscalar mesons, together with the  $\bar{q}q=\bar u u + \bar d d + \bar s s$ created from vacuum, and the quark $s$ from the quark $c$ decay, together with the $ud$ pair of the initial $\Lambda^+_c$, will form the baryon $\Lambda$.}
\label{Fig:mmproduction}
\end{center}
\end{figure}

2) Hadronization: the $u \bar d$ pair from the $W^+$ decay will hadronize into $\pi^+$, and the quark $s$ from the $c$ decay, and the $ud$ pair of the initial $\Lambda^+_c$, together with the $\bar{q}q=\bar u u + \bar d d + \bar s s$ created from vacuum, hadronize into a pseudoscalar meson and a baryon,  as depicted in Fig.~\ref{Fig:mbproduction}. In addition, the $u\bar{d}$ pair from the $W^+$ decay could hadronize into a pair of pseudoscalar mesons, together with the  $\bar{q}q=\bar u u + \bar d d + \bar s s$ created from vacuum, and the quark $s$ from the quark $c$ decay, together with the $ud$ pair of the initial $\Lambda^+_c$, will form the baryon $\Lambda$, as shown in Fig.~\ref{Fig:mmproduction}.

\begin{figure}[htbp]
\begin{center}
\subfigure[]{
\scalebox{0.8}{\includegraphics{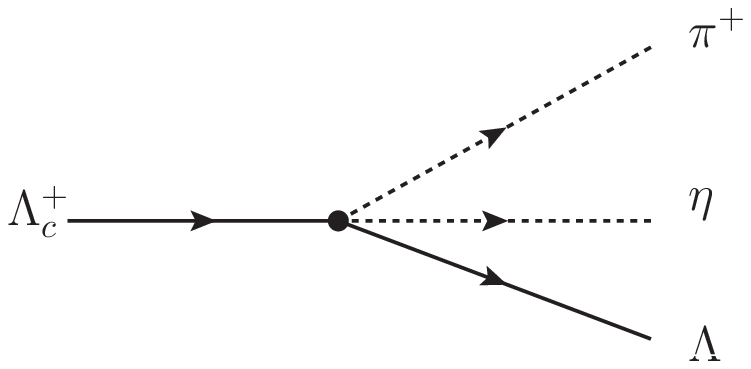}}}
\subfigure[]{
\scalebox{0.8}{\includegraphics{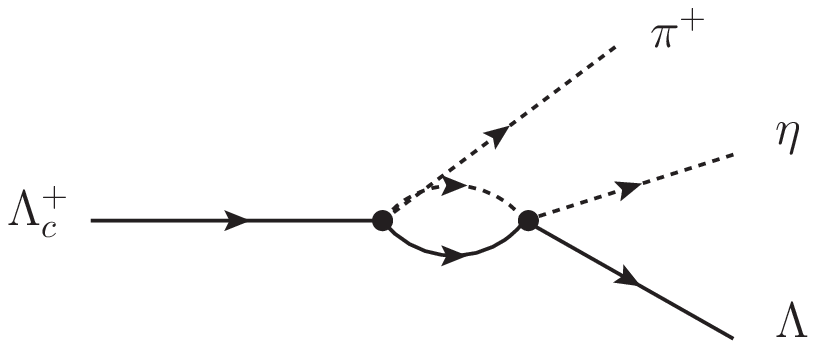}}}
\subfigure[]{
\scalebox{0.8}{\includegraphics{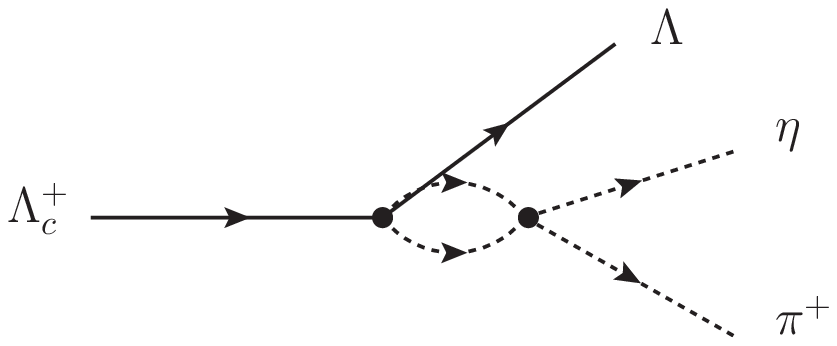}}}
\end{center}
\caption{Mechanism for tree diagram (a), the meson-baryon final-state interaction (b) and the meson-meson final-state interaction (c) for the process $\Lambda_c^+ \to  \eta \Lambda\pi^+$.} \label{Fig:fsi}
\end{figure}

3) Final-state interactions: within the chiral unitary approach, the $S$-wave interactions of the  meson-baryon and meson-meson pairs will  dynamically generate the
$\Lambda(1670)$ and $a_0(980)$, respectively, as shown in Fig.~\ref{Fig:fsi}.

According to Eq.~(6) of Ref.~\cite{Xie:2016evi}, we can write the meson-baryon interaction amplitude of $\Lambda_c^+ \to  \eta \Lambda\pi^+$ decay as follows,
\begin{eqnarray}
T^{MB} &=& V_P \left[ -\frac{\sqrt{2}}{3} + G_{K^- p} (M_{\eta
\Lambda}) t_{K^- p \to \eta \Lambda} (M_{\eta \Lambda})  \right.\nonumber \\
&& + G_{\bar{K}^0 n} (M_{\eta \Lambda}) t_{\bar{K}^0 n \to \eta
\Lambda} (M_{\eta \Lambda}) \nonumber \\
&& \left. -\frac{\sqrt{2}}{3} G_{\eta \Lambda} (M_{\eta \Lambda}) t_{\eta
\Lambda \to \eta \Lambda} (M_{\eta \Lambda}) \right],   \label{Eq:tmb}
\end{eqnarray}
where $V_P$ accounts for the weak decay and hadronization strength, and is
assumed to be independent of the final state interaction. In the
above equation, $G_{MB}$ denotes the one-meson-one-baryon loop
function, which depends on the invariant mass of the  $\eta
\Lambda$ system, $M_{\eta \Lambda}$. The meson-baryon scattering
amplitudes $t_{MB \to \eta \Lambda}$ are obtained in the chiral
unitary approach, which also depends  on $M_{\eta \Lambda}$, and the details
can be found in Refs.~\cite{Oset:2001cn,Oller:2000fj}.

According to Eq.~(13) of Ref.~\cite{Xie:2016evi}, the meson-meson transition amplitude is given as follows,
\begin{eqnarray}
T^{MM} &=& V'_P \frac{2\sqrt{2}}{3} \left[ 1 +  
G_{\pi^+ \eta} (M_{\pi^+ \eta}) t_{\pi^+ \eta \to \pi^+ \eta}
(M_{\pi^+ \eta}) \right.
\nonumber \\
&&\left.  + \frac{\sqrt{3}}{2}G_{K^+ \bar{K}^0} (M_{\pi^+ \eta}) t_{K^+
\bar{K}^0 \to \pi^+ \eta} (M_{\pi^+ \eta})  \right], \label{Eq:tmm}
\end{eqnarray}
where $G_{M M}$ is the loop function of the two-meson
propagators~\cite{Oller:1997ti} and $t_{MM \to \pi^+ \eta}$ are the
meson-meson scattering amplitudes obtained in
Ref.~\cite{Oller:1997ti}, which depend on the $\pi^+ \eta$ invariant mass
$M_{\pi^+ \eta}$. Here $V'_P$ is the weak and hadronization strength for the mechanism of meson-meson interaction. We will take $V'_P = 0.38 V_P$ which is obtained with the assumption that the amplitudes of Eq.~(\ref{Eq:tmb}) and Eq.~(\ref{Eq:tmm}) give rise to the same width for the process $\Lambda_c^+ \to \eta \Lambda \pi^+$ (see more details in Ref.~\cite{Xie:2016evi}).

\subsection{Contribution of  $\Sigma(1385)^+$}\label{sec:1385}

\begin{figure}[htbp]
\begin{center}
\includegraphics[scale=0.6]{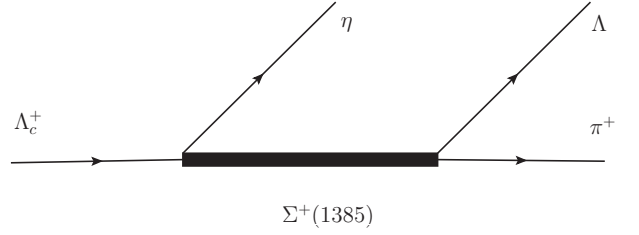}
\caption{Feynman diagram for the contribution of $\Sigma(1385)^+$ to
the $\Lambda_c^+ \to  \eta \Lambda\pi^+$ decay.} \label{Fig:s1385}
\end{center}
\end{figure}

For the contribution of $\Sigma(1385)^+$ in $\Lambda_c^+ \rightarrow \eta\Lambda  \pi^+$, the Feynman diagram is shown in Fig.~\ref{Fig:s1385}. In the non-relativistic reduction~\cite{Lu:2016roh,Pavao:2018wdf,Wang:2015pcn}, the invariant decay amplitude of Fig.~\ref{Fig:s1385} can be written as follows\footnote{In this part we use $\Sigma^*$ represent $\Sigma(1385)$.},
\begin{eqnarray}
T^{\Sigma^{\ast}} 
&=& V''_P \frac{\left|\vec{p}_{\pi}\right|\cdot \left|\vec{p}_{\eta}\right|\cdot\cos\theta}{M_{\pi\Lambda}-M_{\Sigma^{\ast}}+i\frac{\Gamma_{\Sigma^{\ast}}}{2}},
\end{eqnarray}
where $V''_p$ is the relative strength of the contribution of $\Sigma(1385)^+$, while $\vec{p}_{\pi}$ and $\vec{p}_{\eta}$ are the momenta of $\pi^+$ and $\eta$ in the $\pi \Lambda$ rest frame, respectively, and $\theta$ is the angle between $\pi^+$ and $\eta$ in the center of mass frame of the $\pi^+\Lambda$ system~\cite{Wang:2015pcn}, which are given by
\begin{eqnarray}
\left|\vec{p}_{\pi}\right| &=& \frac{\lambda^{\frac{1}{2}}({M_{\pi\Lambda}^{2}},m_{\pi^+}^{2},M_{\Lambda}^{2})}{2M_{\pi\Lambda}}, \\
\left|\vec{p}_{\eta}\right| &=& \frac{\lambda^{\frac{1}{2}}({M_{\Lambda_{c}}^{2}},m_{\eta}^{2},M_{\pi\Lambda}^{2})}{2M_{\pi\Lambda}}, \\
\cos\theta &=& \frac{M^{2}_{\eta\Lambda}-M^{2}_{\Lambda_c}-m_{\pi^+}^{2}+2P^{0}_{\Lambda_c}P_{\pi}^{0}}{2\left|\vec{p}_{\pi}\right|\left|\vec{p}_{\eta}\right|},
\end{eqnarray}
with $\lambda(x,y,z)=x^2+y^2+z^2-2xy-2yz-2xz$. In the $\pi\Lambda$ rest frame, $\vec{p}_{\Lambda_c}=\vec{p}_\eta$, $\vec{p}_\pi=-\vec{p}_\Lambda$, and the $\Lambda_c$ and $\pi$ energies are,
\begin{eqnarray}
&&P^{0}_{\Lambda_c}=\sqrt{M^2_{\Lambda_c}+\left|\vec{p}_{\Lambda_{c}}\right|^2}
 =\sqrt{M^2_{\Lambda_c}+\left|\vec{p}_{\eta}\right|^2},
\nonumber \\
&&P^{0}_{\pi}=\sqrt{m^2_{\pi^+}+\left|\vec{p}_{\pi}\right|^2}.\nonumber
\end{eqnarray}

\subsection{Invariant mass distributions of the $\Lambda_c^+ \to  \eta \Lambda\pi^+$ decay}\label{sec:width}

With all the ingredients obtained in the previous section, one can
write down the modulus squared of the total amplitude as~\cite{Xie:2016evi},
\begin{eqnarray}
|T^{\rm total}|^2=|T^{\rm MB}+T^{\rm MM}|^2+|T^{{\Sigma}^{\ast}}|^2.\label{eq:total}
\end{eqnarray}

The invariant mass distributions for the $\Lambda_c^+ \to \eta \Lambda\pi^+$ decay can be written as~\cite{ParticleDataGroup:2020ssz},
\begin{eqnarray}
\frac{d^{2}\Gamma_{\Lambda^+_c \to  \eta \Lambda\pi^+}}{d{M^{2}_{\pi^+\eta}}d{M^{2}_{\eta\Lambda}}}
&=&\frac{1}{(2\pi)^{3}}\frac{M_{\Lambda}}{8{M^{2}_{\Lambda_{c}^{+}}}}
|T^{\rm total}|^{2}. \label{eq:dgdm1}
\end{eqnarray}
Then the $\pi^+ \eta$ and $\eta \Lambda$ mass distributions can be obtained by integrating over the other invariant mass in Eq.~(\ref{eq:dgdm1}), and the $\pi^+\Lambda$ mass distributions can be obtained by  substituting $M_{\eta \Lambda}$ with $M_{\pi^+\Lambda}$ in Eq.~(\ref{eq:dgdm1}).
For a given value of  $M_{12}$, the range of $M_{23}^2$ is defined as~\cite{ParticleDataGroup:2020ssz}~\footnote{In this work, we label 1, 2, and 3 for final state $\pi^+$, $\eta$, or $\Lambda$.},
\begin{eqnarray}\label{daliz}
(M^2_{23})_{\rm max}=(E_{2}^*+E^*_3)^2-\left(\sqrt{{E_{2}^{*2}}-m^2_{2}}-\sqrt{E^{*2}_{3}-m^2_{3}}\right)^2,\nonumber\\
(M^2_{23})_{\rm min}=(E_{2}^*+E^*_3)^2-\left(\sqrt{{E_{2}^{*2}}-m_{2}^2}+\sqrt{E_{3}^{*2}-m_{3}^2}\right)^2,\nonumber \\
\end{eqnarray}
where  $E_{2}^*$ and $E_{3}^*$ are the energies of particles 2 and 3 in the rest frame of particles 1 and 2, respectively,
\begin{eqnarray}
E_2^*=\frac{M_{12}^2-m_1^2+m_2^2}{2M_{12}}, \nonumber \\
E_3^*=\frac{M_{\Lambda_c}^2- M_{12}^2-m_3^2}{2M_{\Lambda_c}},\nonumber
\end{eqnarray}
with $m_1$, $m_2$ and $m_3$ are the masses of particles 1, 2, and 3, respectively.
All the masses and widths of the particles are taken from the Review of Particle Physics~\cite{ParticleDataGroup:2020ssz}.

\section{Numerical results and discussions} \label{Sec:Results}

With the above formalism, we calculate the invariant mass distributions of the process $\Lambda_c^+ \to  \eta \Lambda\pi^+$. There are three free parameters to be obtained by fitting to the experimental data,
1) $V_p$ for the weak and hadronization strength related to meson-baryon interaction of Fig.~\ref{Fig:mmproduction};
2) $V''_{p}$ for the weight of the contribution of $\Sigma(1385)$;
3) $a_{K\Xi}$ for the subtraction constant in the $K\Xi$ channel of Ref.~\cite{Oset:2001cn}. 
Here we take $a_{K\Xi}$ as a free parameter, because the position of the dynamically generated state $\Lambda(1670)$ is quite sensitive to the value of $a_{K\Xi}$ and only moderately sensitive to $a_{\bar K N}$, $a_{\pi\Sigma}$ and $a_{\eta\Lambda}$, as discussed in Ref.~\cite{Oset:2001cn}. For instance, the pole position is $M_R=1708+i 21$~MeV for $a_{K\Xi}=-2.52$, and $M_R=1680+i 2$~MeV for $a_{K\Xi}=-2.67$ given by Ref.~\cite{Oset:2001cn}. Thus we would like to constrain the pole position of $\Lambda(1670)$ by fitting the parameter $a_{K\Xi}$  to the Belle measurements~\cite{Lee:2020xoz}.

\begin{figure}[htbp]
\begin{center}
\includegraphics[scale=1.]{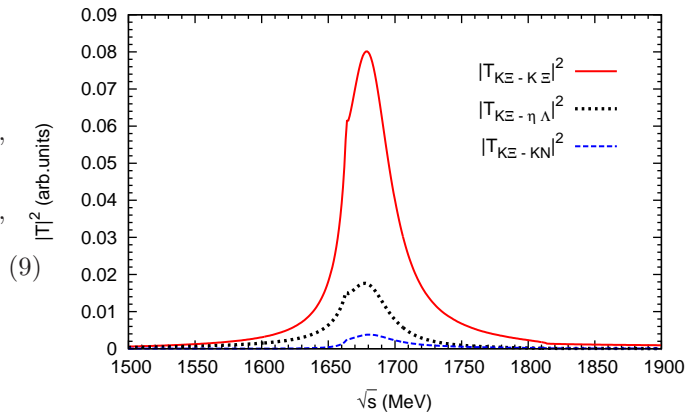}
\caption{(Color online) Modulus squared of the  transition amplitudes $\left|T_{K\Xi-K\Xi}\right|^2$, $\left|T_{K\Xi-\eta\Lambda}\right|^2$, and $\left|T_{K\Xi-KN}\right|^2$, respectively.}\label{fig:t}
\end{center}
\end{figure}

With the model presented above, we perform the fit to the Belle measurements of the $\eta \Lambda$ and $\Lambda \pi^+$ mass distributions~\cite{Lee:2020xoz}. The obtained $\chi^2/d.o.f.$ is $815.4/(91+162-3)=3.26$, and the fitted parameters are: $V_P = 1.32 \pm 0.1$, $V''_P = (3.46 \pm 0.02) \times 10^{-4}$, and $a_{K\Xi} = -2.72 \pm 0.01$. It should be stressed that the obtained $a_{K\Xi}=-2.72$, which is close to the values used in Ref.~\cite{Oset:2001cn}. With this value, we present the modules squared of the transition amplitudes $\left|T_{K\Xi-K\Xi}\right|^2$, $\left|T_{K\Xi-\eta\Lambda}\right|^2$, and $\left|T_{K\Xi-KN}\right|^2$ in Fig.~\ref{fig:t}, where one can find that the position of the pole is about 1680~MeV. 

\begin{figure*}[htbp]
\begin{center}
\includegraphics[scale=1.5]{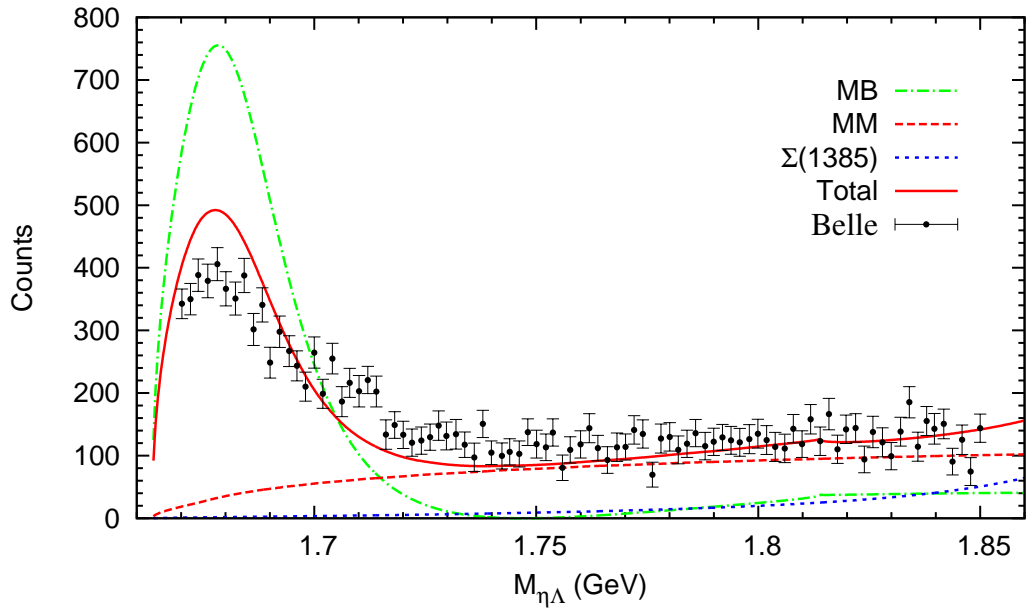}
\caption{(Color online) The  $\eta \Lambda$  invariant  mass distribution
for the $\Lambda_c^+ \to  \eta \Lambda\pi^+$ decay. The curves labeled as `MB', `MM',  and `$\Sigma(1385)$' show the results obtained with $T^{\rm MB}$, $T^{\rm MM}$, and $T^{\Sigma^\ast}$ respectively; `Total' curve corresponds to the total contribution of Eq.~(\ref{eq:total}). The Belle data are taken from Ref.~\cite{Lee:2020xoz}.} \label{fig:etalambda}
\end{center}
\end{figure*}

\begin{figure*}[htbp]
\begin{center}
\includegraphics[scale=1.5]{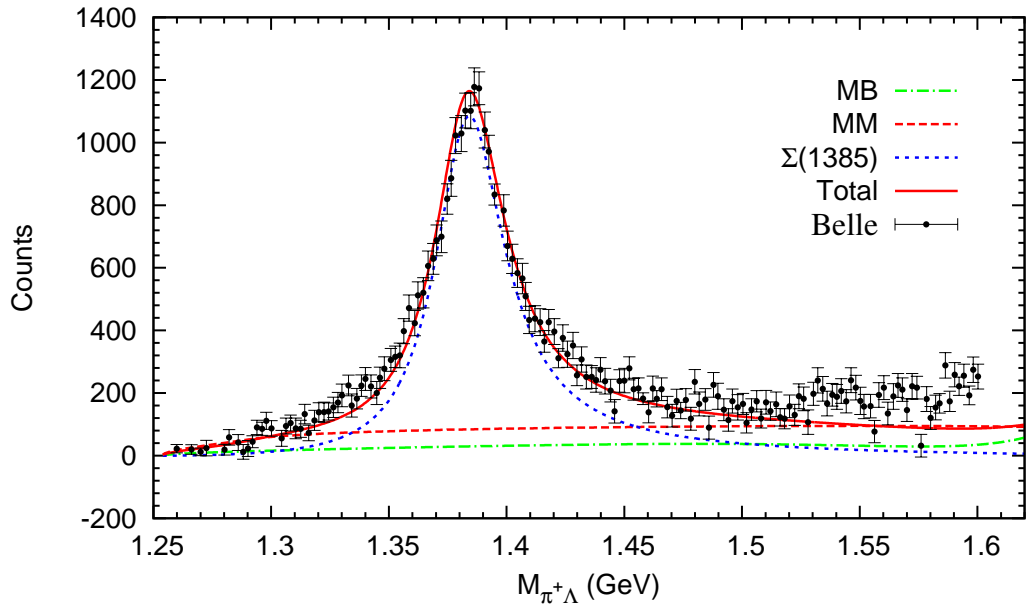}
\caption{(Color online) The  $\pi^+ \Lambda $ invariant mass distribution
for the $\Lambda_c^+ \to  \eta \Lambda\pi^+$ decay. The explanations of the curves are the same as those of Fig.~\ref{fig:etalambda}.}\label{fig:pilambda}
\end{center}
\end{figure*}

With the fitted values of the parameters, we calculate the $\eta \Lambda$ and $ \pi^+\Lambda$ mass distributions, as shown in Figs.~\ref{fig:etalambda} and \ref{fig:pilambda}, respectively, where we show the contribution from meson-baryon interaction (MB) corresponds to $\Lambda(1670)$, meson-meson interaction (MM) corresponds to $a_0(980)$,  $\Sigma(1385)$, the total contribution of all of them (Total), and the Belle measurements (Belle)~\cite{Lee:2020xoz}. For both mass distributions, our results are in good agreement with the Belle measurements. 

From Fig.~\ref{fig:etalambda} one can find that the peak close to the threshold in the $\eta\Lambda$ mass distribution is the signal of the $\Lambda(1670)$, which is dynamically generated from the $S$-wave meson-baryon interaction, and the interference effect between $\Lambda(1670)$ and $a_0(980)$ is  destructive, which is consistent with Ref.~\cite{Xie:2016evi}. The contribution of $\Sigma(1385)$ to the $\Lambda(1670)$ is very small.

In Fig.~\ref{fig:pilambda}, our model calculations can describe well the reported $\pi^+\Lambda$ invariant mass distribution where the $\Sigma(1385)^+$ is clearly shown. It is found that the other contributions is rather small to produce the peak of $\Sigma(1385)$.

\begin{figure*}[htbp]
\begin{center}
\includegraphics[scale=1.5]{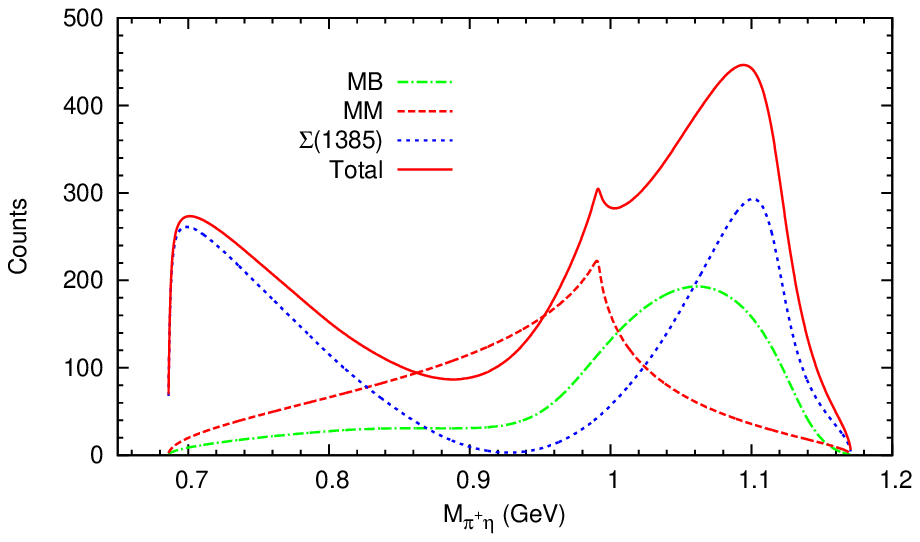}
\caption{(Color online) The   $\eta \pi^+$ invariant mass distribution
for the $\Lambda_c^+ \to \eta \Lambda\pi^+ $ decay. The explanations of the curves are the same as those of Fig.~\ref{fig:etalambda}.}\label{fig:pieta}
\end{center}
\end{figure*}

 Next, we turn to the $\pi^+\eta$ invariant mass distribution. Our numerical results are shown in Fig.~\ref{fig:pieta}. One can find a clear cusp structure around the $K\bar{K}$ threshold, which is consistent with our previous results~\cite{Xie:2016evi}. However, the reflection effect of the $\Sigma(1385)^+$ simultaneously appears in the low energy region and high energy region, which could be easily understood from the Dalitz Plot of the process $\Lambda_c^+ \to  \eta \Lambda\pi^+$ measured by Belle~\cite{Lee:2020xoz}. Thus, the  cusp structure of $a_0(980)$, as one feature of its molecular nature, is expected to be observed in the $\eta\pi^+$ invariant mass distribution.

\section{Summary}

In this work, we have reanalyzed the process of $\Lambda_c^+ \to  \eta \Lambda\pi^+$ by taking into account the $P$-wave contribution from the $\Sigma(1385)$, and the $S$-wave meson-meson and meson-baryon interactions provided by the chiral unitary approach, which dynamically generate the resonances $a_0(980)$ and $\Lambda(1670)$. Our calculated $\eta\Lambda$ and $\Lambda\pi^+$ invariant mass distribution are in good agreement with the Belle measurement, which implies that the peak structure close to the $\eta\Lambda$ threshold in the $\eta\Lambda$ invariant mass distribution could be explained by the dynamically generated resonance $\Lambda(1670)$.

On the other hand, we also show the $\eta\pi^+$ invariant mass distribution, which shows a clear cusp structure of the $a_0(980)$, as one feature of its molecular nature. Due to the $P$-wave coupling of $\Sigma(1385)^+\to  \Lambda\pi^+$, the reflection effect of the $\Sigma(1385)^+$ mainly appears in both the low energy region and high energy region of the $\eta\pi^+$ invariant mass distribution, which could be easily understood from the Dalitz Plot of the process $\Lambda_c^+ \to  \eta \Lambda\pi^+$ measured by Belle~\cite{Lee:2020xoz}.

\section{ACKNOWLEDGMENTS}

This work is partly supported by the China Postdoctoral Science Foundation Funded Project under Grand No.2021M701086, and supported by the Natural Science Foundation of Henan under Grand No.212300410123 and No. 222300420554; the Key Research Projects of Henan Higher Education Institutions under No. 20A140027; the Project of Youth Backbone Teachers of Colleges and Universities of Henan Province (2020GGJS017); the Youth Talent Support Project of Henan (2021HYTP002); the Fundamental Research Cultivation Fund for Young Teachers of Zhengzhou University (JC202041042); the Open Project of Guangxi Key Laboratory of Nuclear Physics and Nuclear Technology, No.NLK2021-08; the National Natural Science Foundation of China under Grants Nos.11735003, 11975041, 11961141004, 12075288, and 11961141012; the Youth Innovation Promotion Association CAS.

\bibliographystyle{ursrt}

\end{document}